\documentclass[aps,prl,twocolumn,showpacs,10pt,superscriptaddress,preprintnumbers,nofootinbib]{revtex4-1}
\usepackage{epsfig,amssymb,amsmath,psfrag,epstopdf,color}
\pdfoutput=1
\usepackage{graphicx}
\usepackage{subcaption}
\usepackage{ragged2e}
\DeclareCaptionJustification{justified}{\justifying}
 \usepackage[normalem]{ulem}
\usepackage{paralist}
\usepackage{hyperref}
\usepackage{xcolor}
\allowdisplaybreaks

\def\beq{\begin{equation}}
\def\eeq{\end{equation}}
\def\bsp#1\esp{\begin{split}#1\end{split}}
\newcommand{\be}{\begin{equation}}
\newcommand{\ee}{\end{equation}}
\newcommand{\bea}{\begin{eqnarray}}
\newcommand{\eea}{\end{eqnarray}}

\def\Fig#1{Fig.~{\ref{#1}}}

\def\to{\rightarrow}

\newcommand{\img}{{\rm i}}
\newcommand{\df}{\mathrm{d}}

\newcommand{\comment}[1]{}

\newcommand{\Tr}{\mathrm{Tr}}

\newcommand{\cE}{\mathcal{E}}

\definecolor{darkgreen}{rgb}{0.13,0.55,0.13}
\definecolor{lightpurple}{rgb}{0.75,0.68,0.88}  

\definecolor{alephblue}{rgb}{0.093473,0.198786,0.543458}

\begin{document}

\preprint{MITP-25-057, MITHIG-MOD-24-001}

\title{Energy Correlators from Partons to Hadrons:\\
Unveiling the Dynamics of the Strong Interactions with Archival ALEPH Data}

\author{Hannah Bossi}
\affiliation{Massachusetts Institute of Technology, Cambridge, MA, 02139}

\author{Yi Chen}
\affiliation{Vanderbilt University, Nashville, TN, 37215, USA}

\author{Yu-Chen Chen}
\affiliation{Massachusetts Institute of Technology, Cambridge, MA, 02139}

\author{Max Jaarsma}
\affiliation{Nikhef, Theory Group, Science Park 105, 1098 XG, Amsterdam, The Netherlands}
\affiliation{Institute for Theoretical Physics Amsterdam and Delta Institute for 
 Theoretical Physics, University of Amsterdam, Science Park 904, 1098 XH Amsterdam, The Netherlands}

\author{Yibei Li}
\affiliation{Mainz Institute for Theoretical Physics, Johannes Gutenberg University, Staudingerweg 9, D-55128 Mainz, Germany}

\author{Jingyu Zhang}
\affiliation{Vanderbilt University, Nashville, TN, 37215, USA}

\author{Ian Moult}
\affiliation{Department of Physics, Yale University, New Haven, CT 06511}

\author{Wouter Waalewijn}
\affiliation{Nikhef, Theory Group, Science Park 105, 1098 XG, Amsterdam, The Netherlands}
\affiliation{Institute for Theoretical Physics Amsterdam and Delta Institute for 
 Theoretical Physics, University of Amsterdam, Science Park 904, 1098 XH Amsterdam, The Netherlands}

\author{Hua Xing Zhu}
\affiliation{School of Physics, Peking University, Beijing 100871, China}
\affiliation{Center for High Energy Physics, Peking University, Beijing 100871, China}

\author{Anthony Badea}
\affiliation{Enrico Fermi Institute, University of Chicago, Chicago IL}

\author{Austin Baty}
\affiliation{University of Illinois Chicago, Chicago, Illinois, USA}

\author{Christopher McGinn}
\affiliation{Massachusetts Institute of Technology, Cambridge, MA, 02139}

\author{Gian Michele Innocenti}
\affiliation{Massachusetts Institute of Technology, Cambridge, MA, 02139}

\author{Marcello Maggi}
\affiliation{INFN Sezione di Bari, Bari, Italy}

\author{Yen-Jie Lee}
\affiliation{Massachusetts Institute of Technology, Cambridge, MA, 02139}


\begin{abstract}
Quantum Chromodynamics (QCD) is a remarkably rich theory exhibiting numerous emergent degrees of freedom, from flux tubes to hadrons. Their description in terms of the underlying quarks and gluons of the QCD Lagrangian remains a central challenge of modern physics.
Colliders offer a unique opportunity to probe these phenomena experimentally: high energy partons produced from the QCD vacuum excite these emergent degrees, imprinting their dynamics in correlations in asymptotic energy flux.
Decoding these correlations requires measurements with exceptional angular resolution, beyond that achieved in previous measurements.
Recent progress has enabled precision calculations of energy flux on charged particles alone, allowing data-theory comparisons for measurements using high resolution tracking detectors.
In this \emph{Letter}, we resurrect thirty-year-old data from the ALEPH tracker, and perform a high angular resolution measurement of the two-point correlation of energy flux, probing QCD over three orders of magnitude in scale in a single measurement.
Our measurement unveils for the first time the full spectrum of the correlator, including light-ray quasi-particle states, flux-tube excitations, and their transitions into confined hadrons. 
We compare our measurement with record precision theoretical predictions, achieving percent level agreement, and revealing interesting new phenomena in the confinement transitions. 
More broadly, we highlight the immense potential of this newly unlocked archival data set, the so called ``recycling frontier", and emphasize synergies with ongoing and future collider experiments.
\end{abstract}

\maketitle


\emph{{\color{alephblue}Introduction.}} The discovery of the Yang-Mills Lagrangian of Quantum Chromodynamics (QCD) as the fundamental theory of the strong interactions serves as one of the great examples of the interplay between theory and experiment in physics. Fifty years after its discovery \cite{Gross:2022hyw}, many of the emergent phenomena arising from the QCD Lagrangian remain only partially understood. Apart from an intrinsic interest in relativistic gauge theories, improving our understanding of QCD is key to advancing diverse areas of physics, ranging from enabling precision studies of the Higgs boson, to understanding the structure of nuclei, and the equations of state of neutron stars.

\begin{figure}
\includegraphics[width=0.45\textwidth]{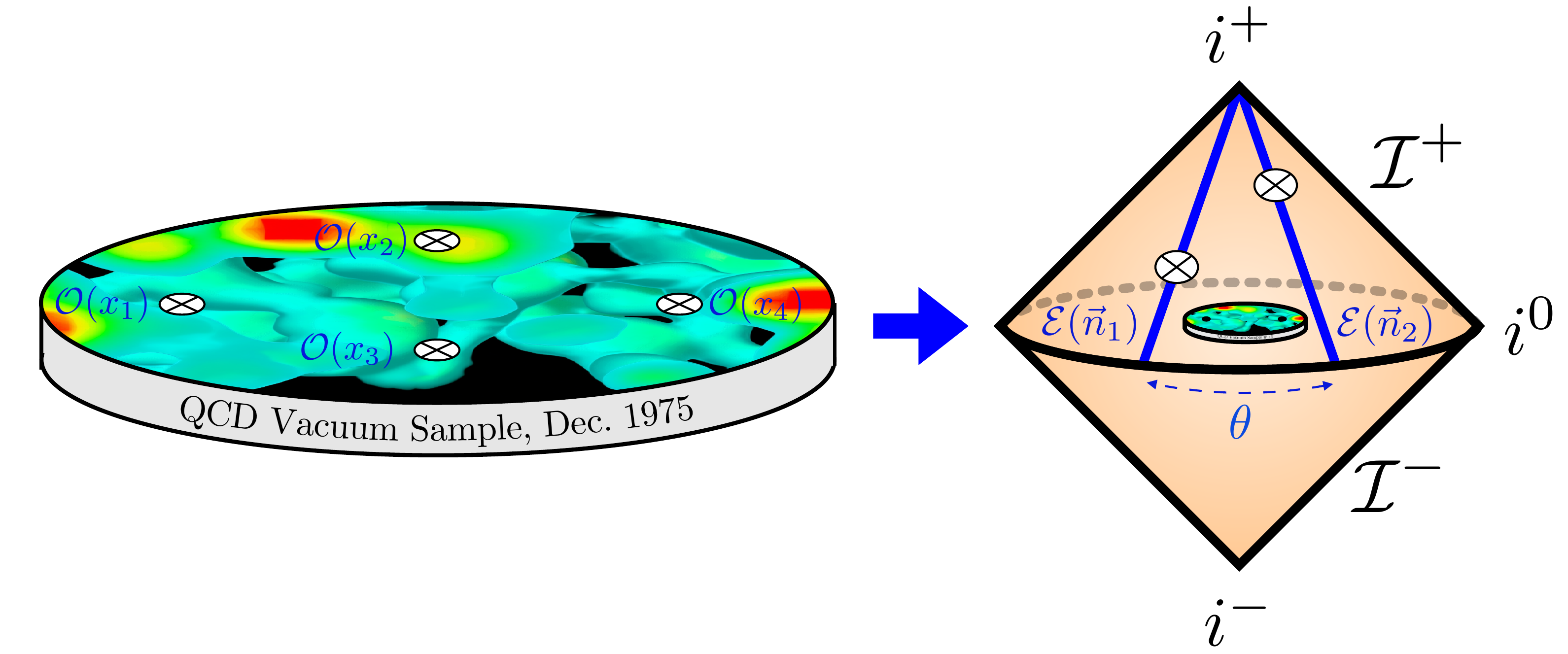}
  \caption{Correlation functions of local operators can be mapped to correlation functions of energy flow operators, $\mathcal{E}(\vec n)$, illustrated as blue lines in the Penrose diagram. These functions characterize asymptotic fluxes, and are directly measurable at $e^+e^-$ colliders.
  }
  \label{fig:1975}
\end{figure}

\begin{figure}
\includegraphics[width=0.4\textwidth]{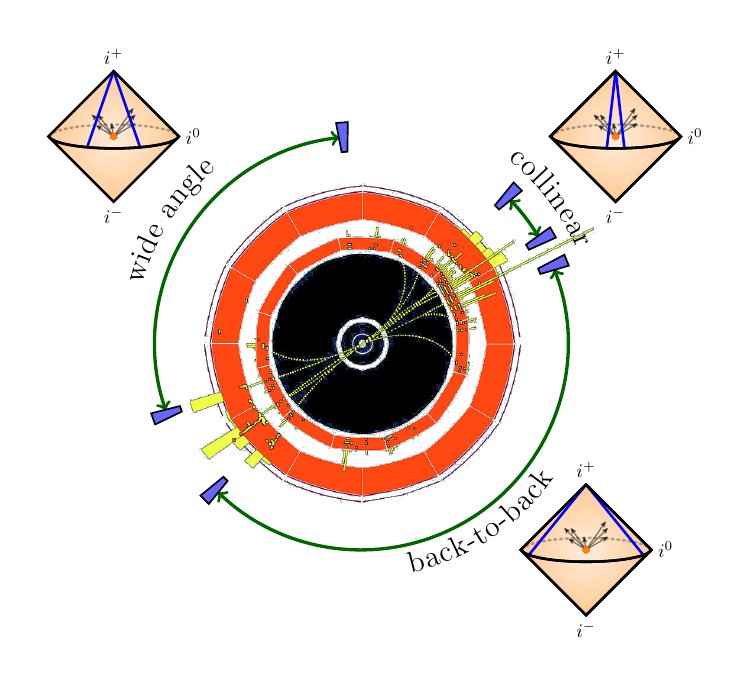}
  \caption{Correlation functions of detector operators can be measured in particle colliders as correlations between pairs of hadrons at different angles. In the center, we illustrate a collision and its corresponding energy depositions in a detector. The Penrose diagrams illustrate the same collision and the corresponding configurations of detector operators.
  }
  \label{fig:detector_plot}
\end{figure}


The complexity of QCD resides in its strong scale dependence, exhibiting a description in terms of asymptotically free quarks and gluons in the ultraviolet (UV), and confined hadronic states in the infrared (IR). At intermediate scales, and in kinematic limits, QCD exhibits numerous additional emergent degrees of freedom, such as flux tubes, Reggeons, and Pomerons. Both their emergence from the underlying quarks and gluons, and their decays into confined hadrons, remain elusive.
While there currently exist no non-perturbative techniques for understanding QCD in Lorentzian signature, there has been tremendous theoretical progress in understanding these phenomena from diverse perspectives, including lattice QCD, effective field theories, and perturbative quantum field theory. However, a complete picture, enabling a description of the transitions between these descriptions, is not yet available.
In light of these theoretical challenges, experimental measurements of these exciting field theoretic phenomena can provide crucial insights. However, this has proven extremely challenging for two reasons: First, emergent states of quarks and gluons cannot be observed directly in experiments, they can only be inferred through their imprint on patterns of hadrons observed in asymptotic detectors. Second, measuring the formation and decay of an emergent state requires measuring observables following the dynamics of QCD over orders of magnitude in scale.
As a result of these combined theoretical and experimental challenges, our understanding of many phenomena in QCD remain murky.

%

Motivated by these long-standing challenges, we present a new and comprehensive program to probe the dynamics of the strong force through ultra-precise measurements of energy–energy correlations in electron–positron collisions. Our approach combines archival data from the Apparatus for LEp PHysics (ALEPH) experiment at the Large Electron Positron Collider (LEP) with a new analysis approach and theoretical advances \cite{Moult:2025nhu} that link the observed correlations to QCD dynamics, transforming legacy datasets into a precision laboratory for strong-interaction physics.

\emph{{\color{alephblue}Imaging States of QCD with Colliders.}} As with any field theory, QCD is characterized by its correlation functions. Electron-positron colliders allow one to directly excite the QCD vacuum with the electromagnetic/electroweak current, $J^\mu$, with a momentum transfer, $q^\mu =(Q,0,0,0)$, controlled by the colliding electrons. This enables a direct measurement of the two-point function (total cross section)
\begin{align}\label{eq:two_point}
\sigma(q^2) &=\int \df^4x\, L_{\mu \nu} e^{\img q\cdot x} \langle0| J^\mu(x) J^\nu(0)|0 \rangle\,,
\end{align}
providing access to the spectral function, and spectrum of hadronic resonances. Here $L_{\mu \nu}$ denotes a particular contraction, referred to as the leptonic tensor.

However, many interesting features of quantum field theory (QFT) are only accessible in higher point correlation functions. Arguably, the most important example are four point functions, illustrated in a sample of the QCD vacuum in \Fig{fig:1975}. Correlation functions of local operators probe the theory at the scale $(x_i-x_j)^2$, which can be either Euclidean or Lorentzian. By tuning the kinematics (positions) of the operators, we are able to probe the theory as a function of scale, selecting the states that dominate the correlator, and cleanly isolating emergent degrees of freedom of QCD. 
Mathematically, this is codified through operator product expansions (OPEs). 
Our understanding of the structure of four-point functions, and the relevant OPEs has advanced tremendously due to progress from the conformal bootstrap \cite{Poland:2018epd}, providing a systematic understanding of their ability to access numerous interesting phenomenon, such as Regge limits, light-cone limits, or high-spin dynamics.  Beyond their intrinsic interest, kinematic limits of correlators provide an opportunity for precision tests of QCD, since they can be computed to extremely high precision.

Unfortunately, multi-point correlation functions of local operators in QCD are not directly measurable at collider experiments, since they would require measuring devices with resolution at the scale of the fundamental interactions: probing quark and gluon degrees of freedom would require a resolution of $\Delta x \sim 10^{-16}$m, or $\Delta t\sim 10^{-27}$s.  Instead, we only observe hadrons in detectors far from the collision. This significantly complicates our ability to access the UV dynamics of the theory, since it must be inferred from patterns imprinted in the IR.

Fortunately, while we cannot measure multi-point correlation functions of local operators, we \emph{can} measure multi-point correlation functions of a class of non-local operators referred to as ``detector operators". The most famous of these is the ``energy flow", or ``average null energy" (ANE) operator, which is expressed as an integral of the stress tensor \cite{Sterman:1975xv,Sveshnikov:1995vi,Tkachov:1995kk,Korchemsky:1999kt}
\begin{align}\label{eq:ANEC_op}
\mathcal{E}(\vec {n}) = \lim_{r\to \infty}  \int\limits_0^\infty \df t\, r^2 n_i T_{0i}(t,r \vec{n})\,.
\end{align}
 The ANE operator appears in many areas of physics, ranging from Hawking's chronology protection conjecture \cite{Hawking:1991nk} in general relativity, to constraints on renormalization group flows \cite{Hartman:2023ccw,Hartman:2023qdn}, thus providing a connection between collider phenomenology,  and the deep underlying principles of QFT.   Unlike a local operator which is specified by a coordinate position, the energy flow operator is specified by a unit vector, $\vec{n}$. It is convenient to illustrate energy flow operators using Penrose diagrams, as in \Fig{fig:1975}, where they are depicted as lines at specific angular positions at future null infinity ($\mathcal{I}^+$), emphasizing the integral over time appearing in their definition.

Energy flow operators establish a non-perturbative relation between the distributions of hadrons observed in $e^+e^-$ collisions, and correlation functions of stress tensor operators in QCD. In particular, the four-point correlator of two current and two stress tensors, $\langle0| J(x) T(y_1) T(y_2) J(0)|0 \rangle$, is related to the two-point function of energy flow operators, the so-called ``energy-energy correlator" (EEC)\cite{Basham:1979gh,Basham:1978zq,Basham:1978bw,Basham:1977iq}
\begin{align}\label{eq:EEC_op_def}
\hspace{-0.35cm}\text{EEC}&=
 \int \df^4x\, \frac{e^{\img q\cdot x}}{\sigma Q^2} L_{\mu \nu} \langle0| J^\mu(x) \cE(\vec{n}_1) \cE(\vec{n}_2) J^\nu(0)|0 \rangle 
\,.
\end{align}
After averaging over the polarizations of the current, the two-point energy correlator is a function of $z=(n_1\cdot n_2)/2=(1-\cos(\theta))/2\,,$ with $\theta$ the physical angle between the two energy flow operators, as illustrated in \Fig{fig:1975}.

Correlation functions of energy flow operators can be measured experimentally by using their action on asymptotic states \cite{Sterman:1975xv},\footnote{This definition was introduced 50 years ago, enabling the theoretical study of energy flux in Yang-Mills theories, and motivating our labeling of the sample in \Fig{fig:1975}.} 
\begin{align}
\mathcal{E}(\vec{n}) |X\rangle= \sum_i k_i^0 \delta(\Omega_{\vec{n}} -\Omega_{\vec{k}_i})|X\rangle\,,
\end{align}
to express the  EEC as an energy weighted cross section, 
\begin{align}
  \label{eq:EEC_def}
  \text{EEC}(z)= \sum_{i,j}\int d\sigma\ \frac{E_i E_j}{Q^2} \delta\left(z - \frac{1 - \cos\theta_{ij}}{2}\right) \,.
\end{align}
 Here $E_i$ denotes the energy of particle $i$, and $\theta_{ij}$ the angle between a pair of particles. 
 This relation is illustrated in Fig. \ref{fig:detector_plot}, where we highlight both the physical picture of the collider, showing correlations being measured between pairs of particles at different angles, as well as Penrose diagrams highlighting the corresponding positions of the detector operators in spacetime. 
This definition was used to measure the energy correlator in early $e^+e^-$ colliders, including  PLUTO \cite{PLUTO:1985yzc,PLUTO:1979vfu},
CELLO \cite{CELLO:1982rca}, JADE \cite{JADE:1984taa}, MAC \cite{Fernandez:1984db}, MARKII \cite{Wood:1987uf}, TASSO \cite{TASSO:1987mcs}, AMY \cite{AMY:1988yrv}, TOPAZ\cite{TOPAZ:1989yod}, ALEPH\cite{ALEPH:1990vew}, L3 \cite{L3:1991qlf,L3:1992btq}, DELPHI \cite{DELPHI:1990sof}, OPAL \cite{OPAL:1990reb,OPAL:1991uui}, and SLD \cite{SLD:1994idb}. These measurements typically focused on the bulk angular region of the correlator ($z\sim 1/2$) to extract the value of $\alpha_{s}$ from experimental data. The definition provided here has also been extended to the study of two-point \cite{Komiske:2022enw,CMS:2024mlf,ALICE:2024dfl,ALICE:2025igw,Tamis:2023guc,STAR:2025jut,CMS-PAS-HIN-23-004,CMS:2025jam,CMS:2025ydi} and three-point \cite{Komiske:2022enw,Chen:2022swd} energy correlators inside high energy jets at hadron colliders.

While correlation functions of local operators probe the theory at the scale $(x_i-x_j)^2$, correlation functions of energy flow operators probe the dynamics of the theory at the scale of the \emph{angular} separations of the detector operators. For two-point energy correlator, this corresponds to the scales $\mu_z^2=Q^2z$ and $\mu_{(1-z)}^2=Q^2(1-z)$. By performing measurements as a function of the angle, this provides the remarkable opportunity to measure the dynamics of QCD over orders of magnitude in a single well controlled observable, and at a single collider with \emph{fixed} energy. At high energy colliders, we can probe QCD from the scale $\mu^2=Q^2$, where it is described by weakly interacting partons, all the way to below the confinement scale, $\mu_z^2< \Lambda_{\text{QCD}}^2$, or $\mu_{(1-z)}^2< \Lambda_{\text{QCD}}^2$, giving experimental access to the confinement transition, and all the intermediate dynamics.
Achieving this in reality requires high angular resolution measurements in the limits when the detectors are brought collinear ($z\to 0$), and back-to-back ($z\to 1$), as illustrated in \Fig{fig:detector_plot}. This angular resolution was not achieved in earlier measurements, which furthermore did not focus on these kinematic limits. Recent measurements at hadron colliders have studied the collinear limit \cite{Komiske:2022enw,CMS:2024mlf,ALICE:2024dfl,ALICE:2025igw,Tamis:2023guc,STAR:2025jut,CMS-PAS-HIN-23-004,CMS:2025jam,CMS:2025ydi}, however, precision measurements of the entire spectrum, in particular, the back-to-back limit, have never been performed.

\begin{figure}
\includegraphics[width=0.45\textwidth]{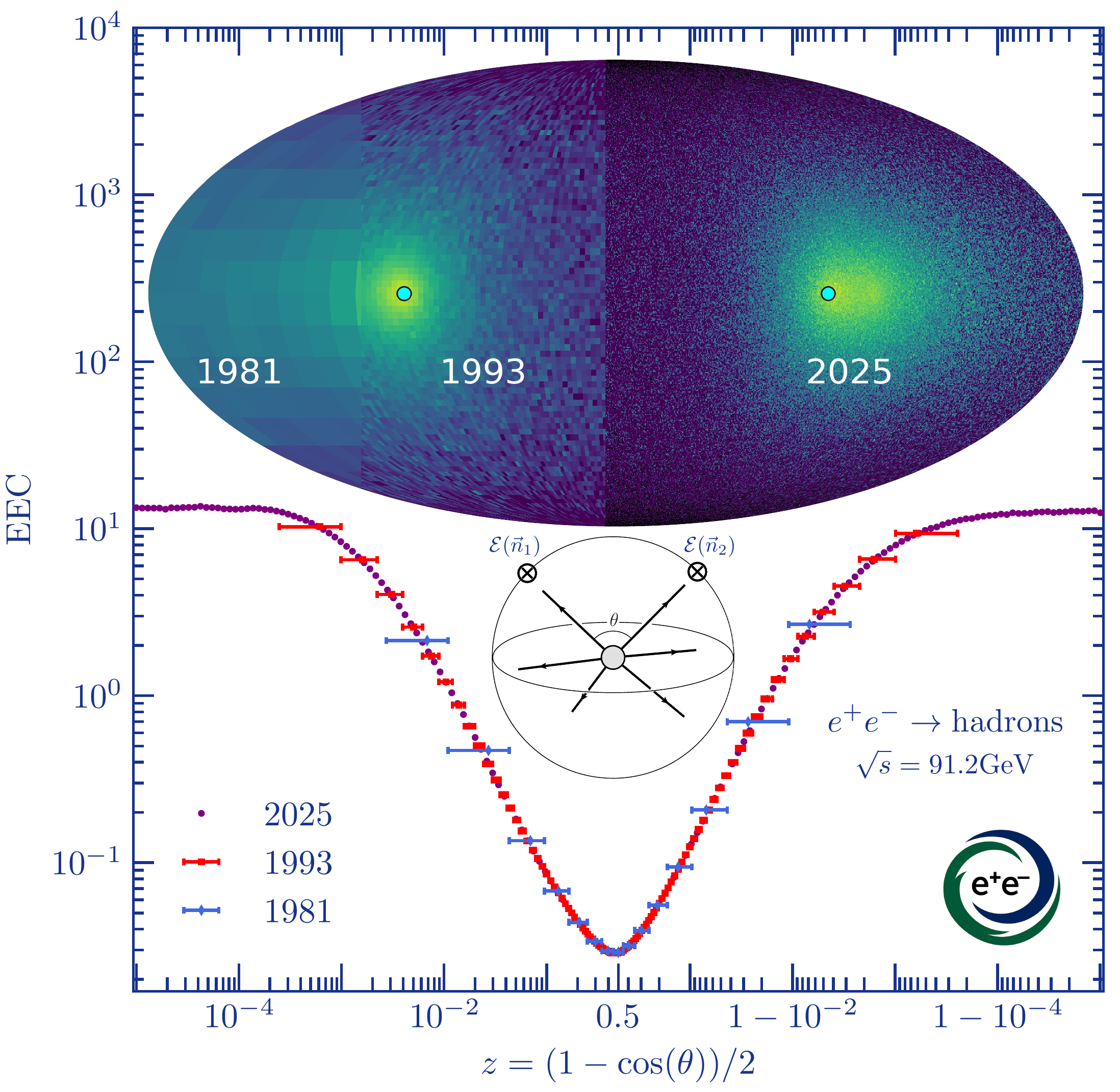}
  \caption{The energy flux in $e^{+}e^{-}$ collisions, and the corresponding EEC distribution, viewed with three angular resolutions: the first PLUTO measurement (1981), LEP measurements with hadronic calorimeters (1993), and our track based ALEPH measurement (2025). The two energy flow operators are illustrated as green dots.
}
  \label{fig:CMB_plot}
\end{figure}

\begin{figure*}
\includegraphics[width=0.95\textwidth]{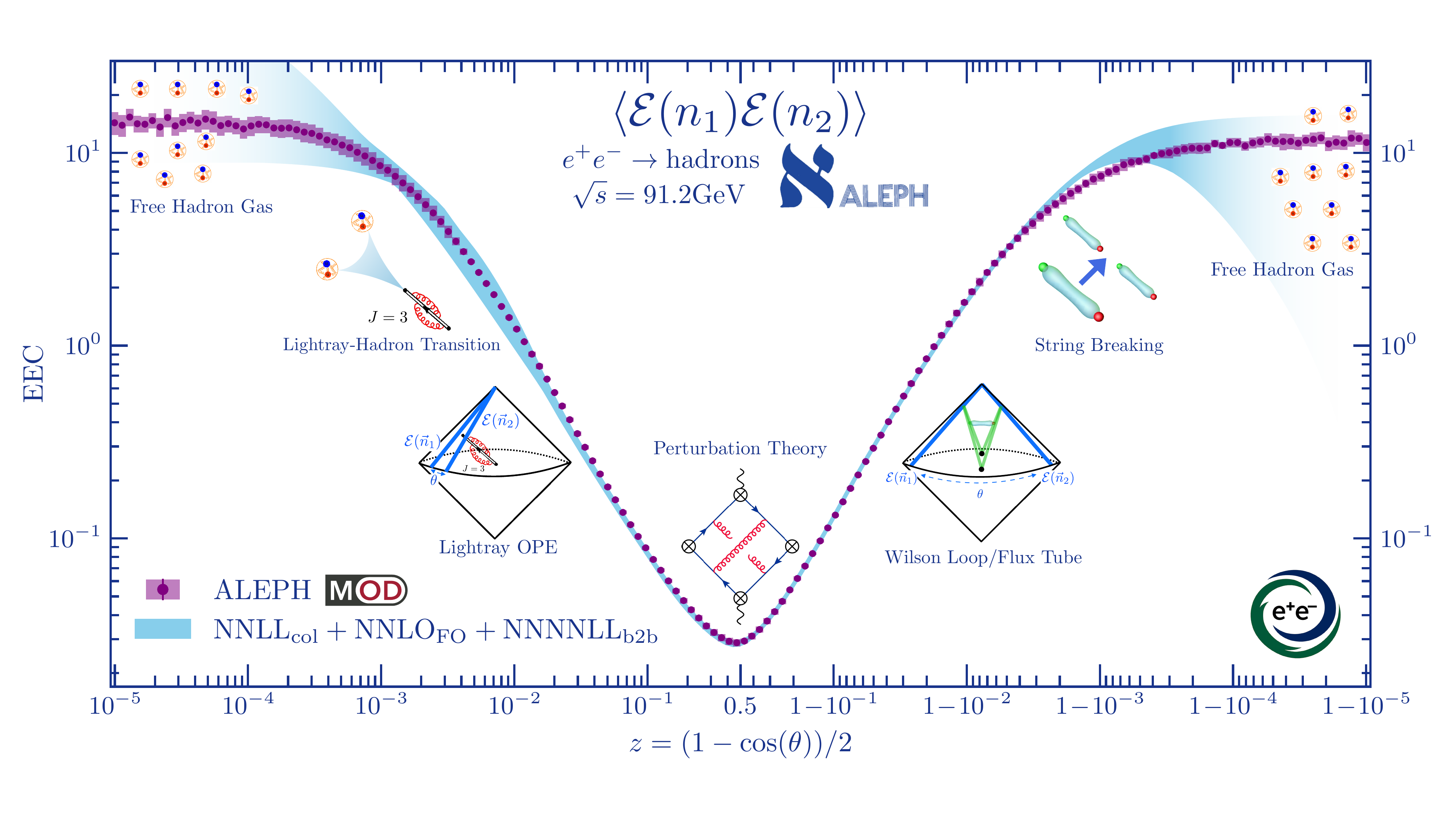}
  \caption{The energy correlator from partons to hadrons: A measurement of the EEC using archival data from the ALEPH tracker. Statistical error bars are shown as vertical lines and systematic error bars as purple boxes. ALEPH data is compared with state-of-the-art theoretical predictions, shown in light blue. Our re-analysis greatly extends both the precision, and angular range of the measurement, enabling a complete view of the dynamics of QCD.
  }
  \label{fig:plot_overall}
\end{figure*}

\emph{{\color{alephblue}Sharpening the Collider Lens with Tracking.}} Achieving a measurement of the full spectrum of the EEC, from partons to hadrons, requires achieving an angular resolution of $z\,, (1-z) \ll (\Lambda_{\text{QCD}}/Q)^2$. Unfortunately, the resolution of the LEP detectors, which have been retired for over two decades, cannot be improved, and there are no current high energy $e^+e^-$ colliders. One may therefore be led to the depressing conclusion that the full structure of the EEC will remain obscured until a new $e^+e^-$ collider is constructed.

Fortunately, there is loophole: The angular resolution of previous measurements was not set by a fundamental limitation of the detectors, but rather by the way that the data was analyzed to enable comparisons with theoretical calculations.  Indeed, a fundamental theorem in QFT, the Kinoshita-Lee-Nauenberg \cite{Kinoshita:1962ur,Lee:1964is} theorem, states that an observable can only be computed in \emph{perturbation theory} if it sums over all degenerate states of a fixed energy. This theorem has direct implications for experimental analyses that wish to compare with precision perturbative calculations, since it necessitates the measurement of both neutral and charged particles when constructing the correlator. Charged particles are measured by the tracking detector, which has an exceptional angular resolution. However, neutral particles are measured by the hadronic calorimeter, which despite having excellent energy resolution, has a worse angular resolution compared to the tracker, ultimately limiting the angular resolution of energy correlator measurements.  

Calculations on tracks are challenging, since they require theoretical control from the scale of the collision, $Q\sim 90$ GeV, to the scale of hadrons, $\Lambda\sim 1$ GeV. Driven by improvements in our understanding of effective field theories, and the application of renormalization group techniques to QCD, it has become possible to perform systematically improvable calculations of energy correlator observables on tracks \cite{Chang:2013rca,Chang:2013iba,Jaarsma:2023ell,Chen:2022pdu,Chen:2022muj,Jaarsma:2022kdd,Li:2021zcf}.  This is achieved by performing a matching calculation  
\begin{align}\label{eq:replace_tracks}
\cE_{\text{tr}}(\vec n_1)=T_q(1;\mu) \cE_q (\vec n_1;\mu)+  T_g(1;\mu) \cE_g (\vec n_1;\mu) \,,
\end{align}
between a non-perturbative detector, $\cE_{\text{tr}}(\vec n_1)$, and perturbative detectors $\cE_g (\vec n_1;\mu)$, $\cE_q (\vec n_1;\mu)$, introducing a dependence on the renormalization group scale, $\mu$. The non-perturbative matching coefficients are referred to as track functions, and have been experimentally measured by the ATLAS collaboration \cite{ATLAS:2024jrp}. This modification to the experimental measurement preserves the direct link to the underlying four-point function of the QFT, but significantly enhances the possible resolution which can be achieved in experiment.

The ability to directly compare track based measurements with precision theoretical calculations completely transforms the potential of the LEP data, opening the potential for high angular resolution measurements directly using data from LEP tracking detectors, and effectively providing a new dataset that has been preserved in a time-capsule. Heroic efforts have made this a reality, resurrecting the LEP data from the ALEPH and DELPHI experiments, and enabling it to be re-imagined through a modern lens~\cite{Badea:2019vey,Chen:2021uws,Chen:2023njr,Zhang:2025nlf,Electron-PositronAlliance:2025hze}.   

To illustrate the transformative improvement in angular resolution achievable using the ALEPH tracking detector, in \Fig{fig:CMB_plot} we show an illustration of the energy flux in $e^+e^-$ collisions, viewed with three different angular resolutions corresponding to the first measurement of the energy correlators by PLUTO in 1981, the LEP measurement of the energy correlator in 1993, and the resolution that we will achieve using the ALEPH tracker (labeled 2025). This directly translates into the angular resolution of the energy correlator measurement, which is simulated for different resolutions in \Fig{fig:CMB_plot} (the real data is shown in \Fig{fig:plot_overall}).

\begin{figure}
\includegraphics[width=0.45\textwidth]{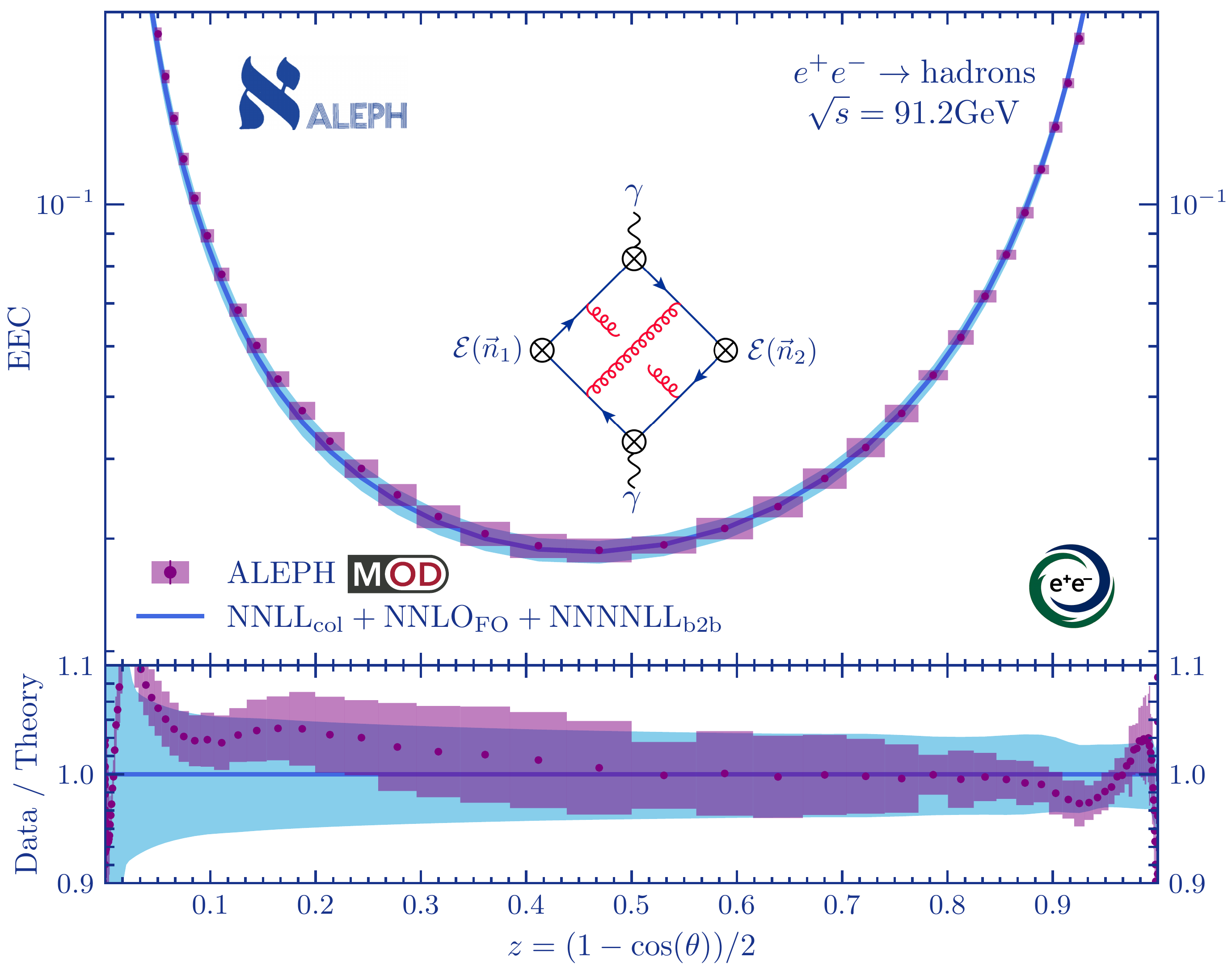}
  \caption{The EEC in the perturbative $z\sim 1/2$ region. 
  }
  \label{fig:bulk}
\end{figure}

\emph{{\color{alephblue}Archival ALEPH Data Analysis.}} In this \emph{Letter}, we use archived data from the ALEPH tracker to perform a high angular resolution measurement of the two-point energy correlator on charged particles, breaking the resolution barrier $z\sim (1-z) \ll (\Lambda_{\text{QCD}}/Q)^2$, and enabling a complete view of the energy correlator from partons to hadrons. Here we provide a brief summary of the analysis. Additional details are provided in the corresponding analysis notes for the ALEPH measurement \cite{Bossi:2025xsi} and a related DELPHI measurement \cite{Zhang:2025nlf}.

Our measurement utilizes archived $e^{+}e^{-}$ annihilation data at a center-of-mass energy of 91.2 GeV that was collected by the ALEPH experiment at LEP in 1994, for which
a dedicated tune of $\textsc{Pythia}$ 6.1~
\cite{Sjostrand:2000wi} Monte Carlo (MC) simulation from the same period is also available. The ALEPH detector features a tracking system composed of  a two-layer silicon strip vertex detector, a cylindrical drift chamber, and a large time projection chamber
(TPC) for the measurement of charged particles. These tracking detectors are immersed in a 1.5 T axial magnetic field generated
by a superconducting solenoidal coil. For this analysis, the archived data and MC samples were reformatted into an MIT Open Data format~\cite{Tripathee:2017ybi}, previously validated through two-particle correlation function studies~\cite{Badea:2019vey}. 

Hadronic events were selected by requiring the polar angle ($\theta_{\text{lab}}$) of the event sphericity axis~\cite{Heister:2003aj} in the laboratory frame to lie between $7\pi/36$ and $29\pi/36$, ensuring full containment within the detector acceptance. To suppress contamination from leptonic processes, events were required to have at least five charged tracks with a combined minimum energy of 15 GeV~\cite{Barate:1996fi}. Under these selection criteria, residual backgrounds such as $e^+e^-\rightarrow\tau^{+}\tau^{-}$ were estimated to contribute less than 0.26\% of the events~\cite{Barate:1996fi}. A total of approximately 1.36 million hadronic events, predominantly originating from $Z^{0}$ boson decays into quark-antiquark pairs, were analyzed. Charged particle tracks were selected using stringent quality criteria identical to those applied in previous ALEPH analyses~\cite{Barate:1996fi}, including a minimum transverse momentum relative to the beam axis ($p_{\rm T}^{\rm lab}$) of 0.2 GeV/$c$ and $|\cos{\theta_{\text{lab}}}| < 0.94$. algorithm~\cite{Barate:1996fi}. 

To confirm the correctness of the data handling and selection procedures, event thrust distributions~\cite{Farhi:1977sg} were compared to previously published results from the ALEPH Collaboration~\cite{Heister:2003aj}. The distributions were successfully reproduced within uncertainties, validating the reliability of the current data analysis framework.

We unfold the detector-level EEC distribution for detector effects using the D’Agostini iterative method~\cite{DAgostini:2010hil}. To characterize bin-to-bin migration from finite track angular and momentum resolutions, a two-dimensional response matrix binned in $z$ and the energy weights of the track pairs, $E_{i}E_{j}$, is constructed. This matrix is built using detector- and generator-level tracks matched via the Hungarian method~\cite{hungarianMatching}. Contributions from unmatched tracks, which are considered mis-reconstructed, are subtracted from the data before unfolding. Bin-by-bin corrections are applied after unfolding to account for inefficiencies from both unmatched tracks and event selections. These correction factors are derived from the archival $\textsc{Pythia}$ 6 MC with Quantum Electrodynamics (QED) effects from initial state radiation removed. Final state QED effects are negligible because the measurement only includes charged tracks, and the probability of a photon converting to an electron-positron pair is small.

We consider a variety of sources of systematic uncertainties, including variations in the number of TPC hits, detector track to generator level particle matching scheme, the energy-weight binning choice, the D'Agostini regularization strength, and the prior dependence of the unfolding.
The dominant uncertainty comes from the prior dependence. As the archived sample is the only ALEPH detector-level simulation currently available, a reweighting approach is used where the two-dimensional EEC distribution from the archival $\textsc{Pythia}$ 6.1 MC is re-weighted to match reconstructed data. Repeating the unfolding procedure with this reweighted MC results in about 5\% in the region around $z=0.5$, which increases to 10\% in the collinear and back-to-back limits. 
All other systematic uncertainties contribute less than 2\% each.

\begin{figure}
\includegraphics[width=0.40\textwidth]{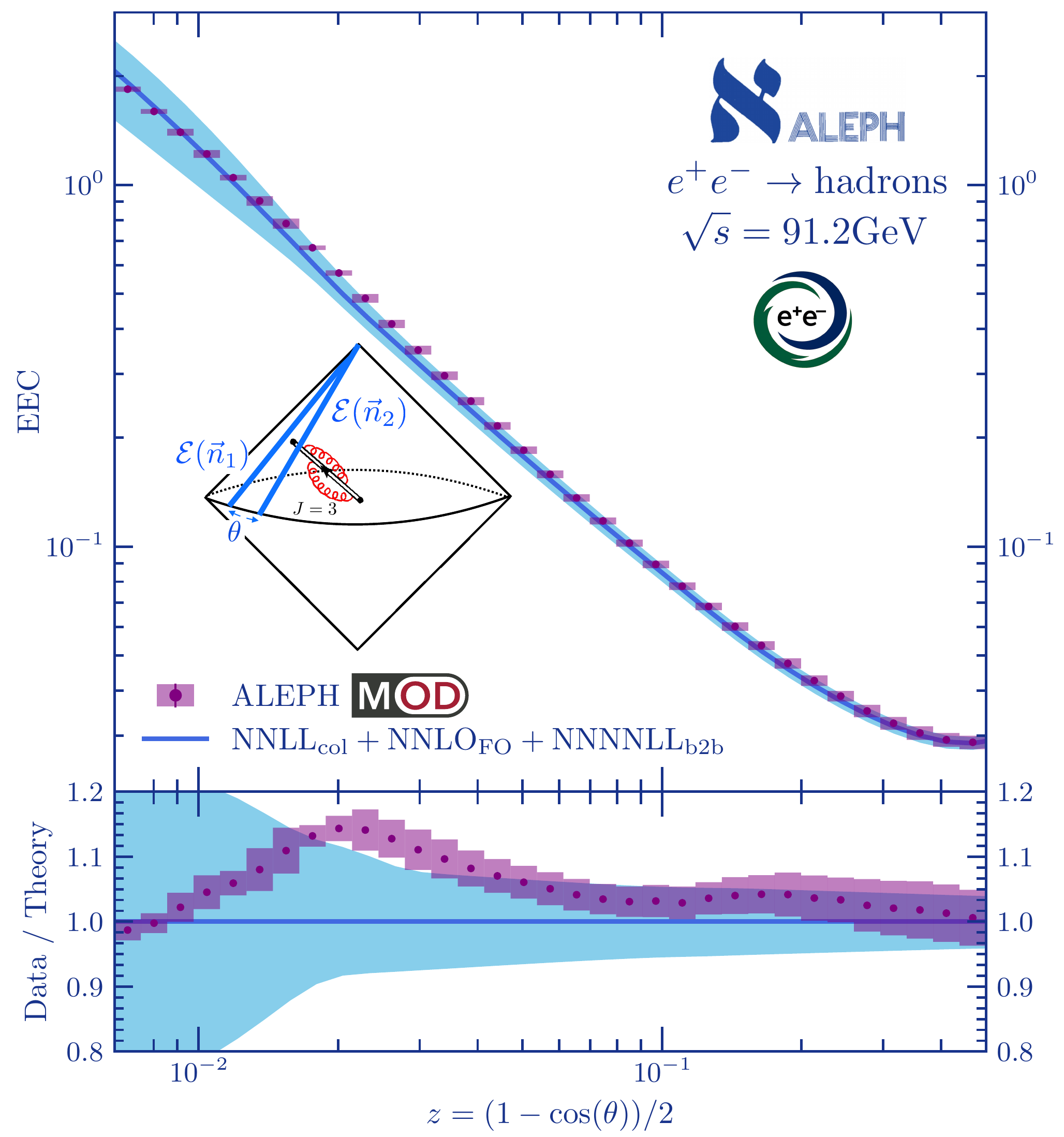}
\includegraphics[width=0.40\textwidth]{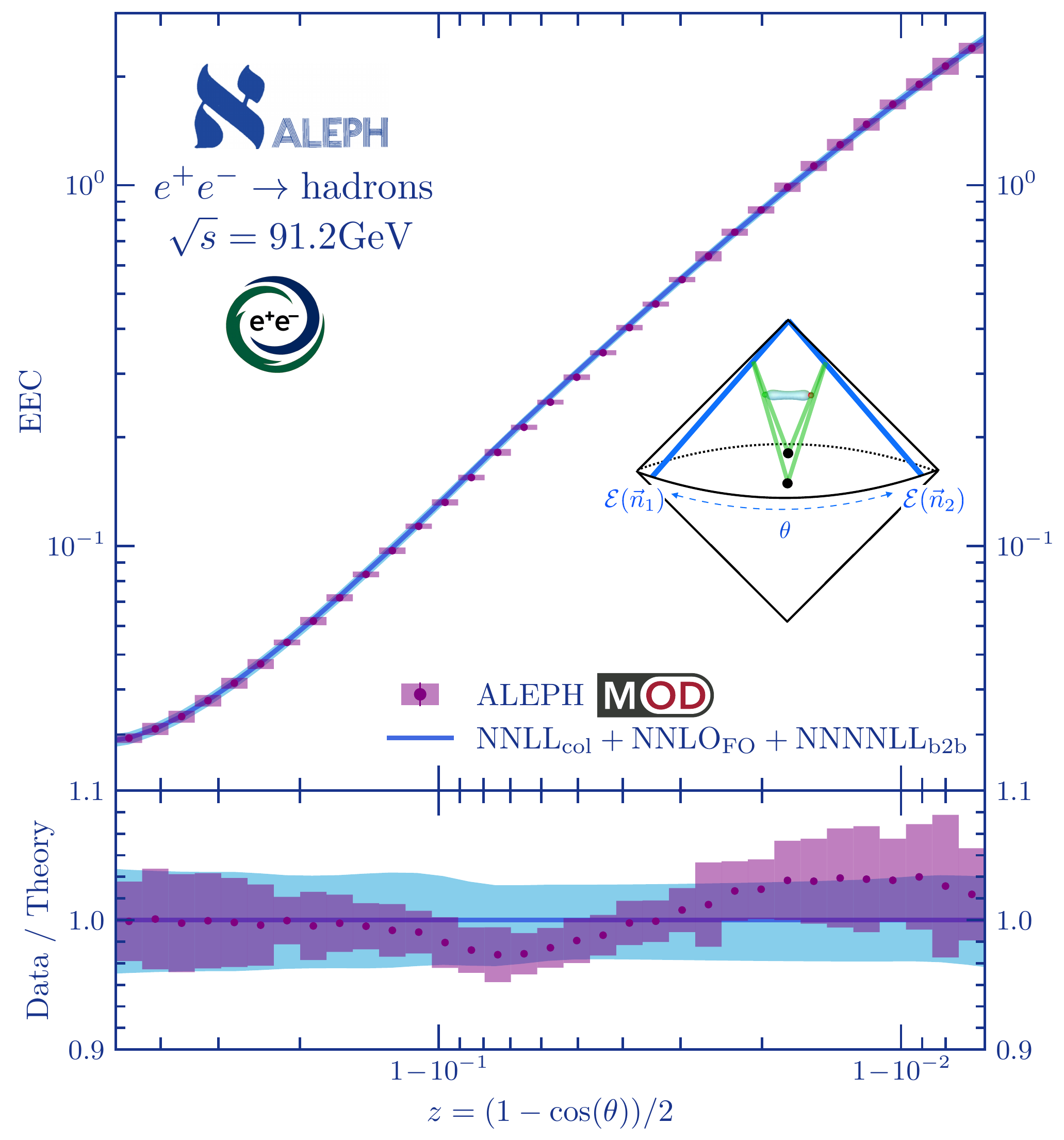}
  \caption{The EEC in the perturbative collinear ($z\to 0$), and back-to-back ($z\to 1$) limits, illustrating a precision understanding of its kinematic limits. 
  }
  \label{fig:pert}
\end{figure}

\begin{figure}
\includegraphics[width=0.40\textwidth]{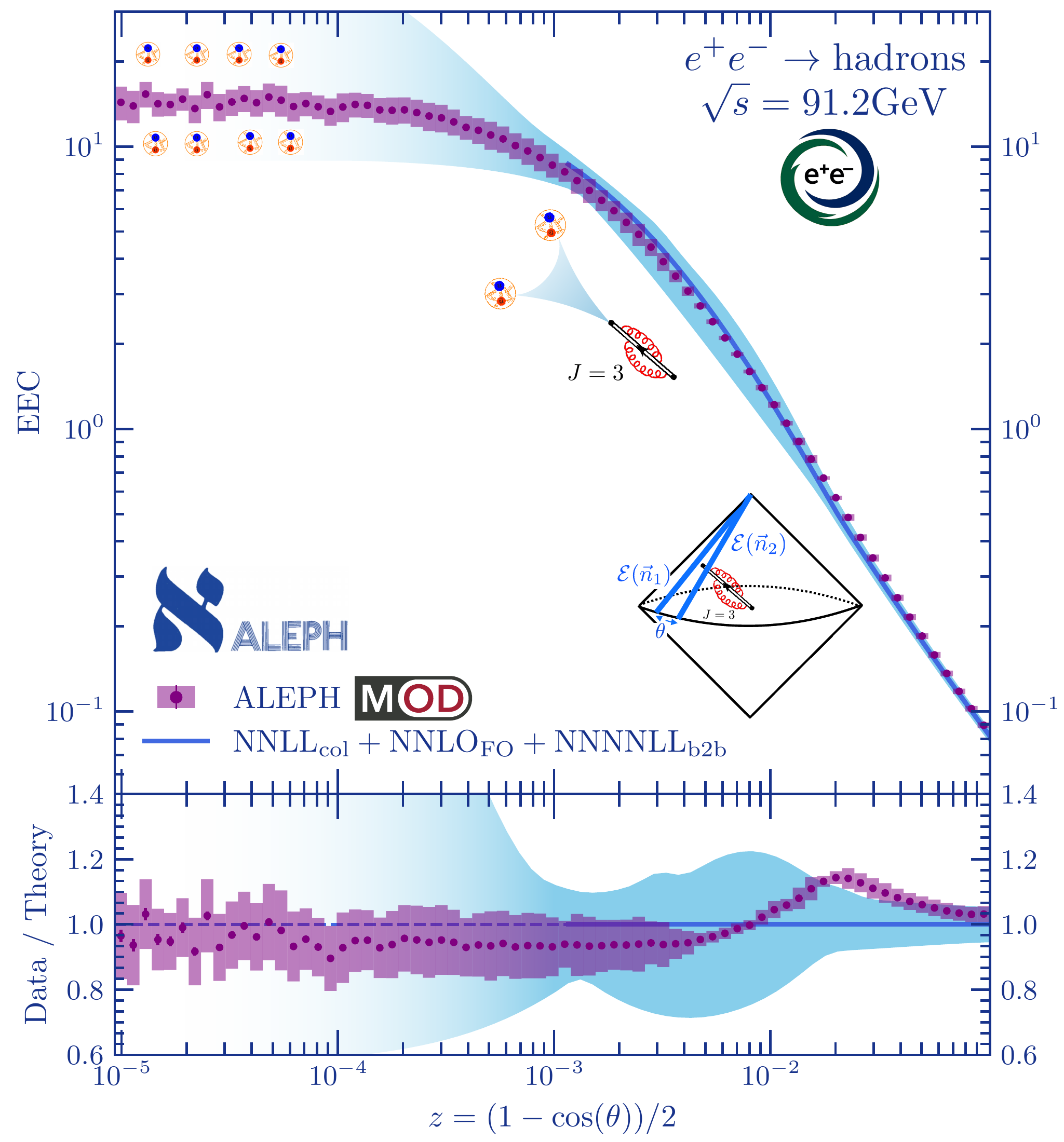}
\includegraphics[width=0.40\textwidth]{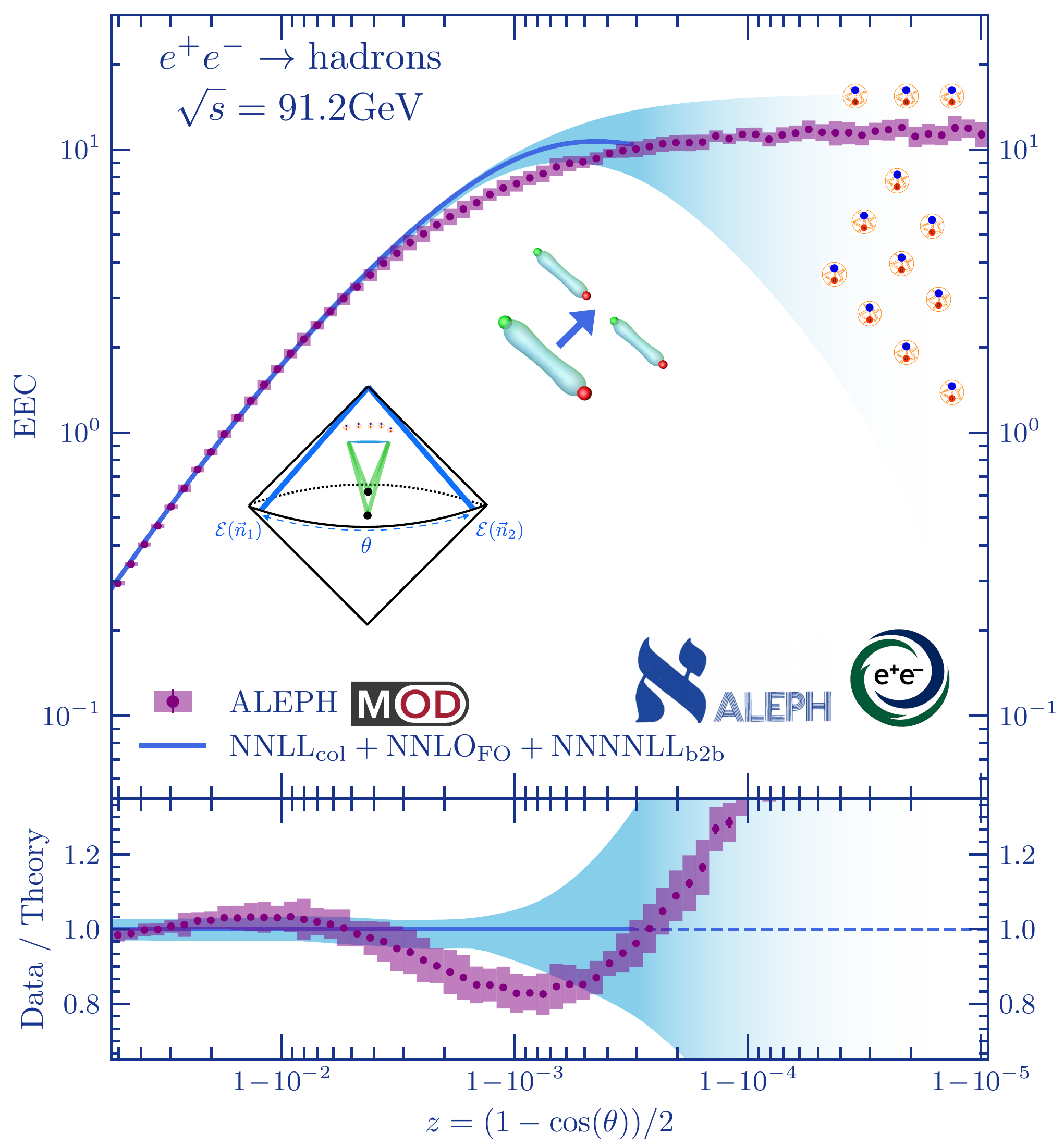}
  \caption{Precision measurements of the confinement transition in the collinear and back-to-back limits of the EEC. 
  }
  \label{fig:transitions}
\end{figure}

The fully corrected spectra are shown in \Fig{fig:plot_overall}, which cover five orders of magnitude in $z$. To illustrate the exceptional angular resolution, we use a logarithmic scale in $(1-z)$ in the back-to-back limit, and a logarithmic scale in $z$ in the collinear limit.  Statistical error bars are displayed as vertical lines, and systematic error bars as boxes. This provides the sharpest ever view of energy flux in QCD, unveiling a remarkable structure, highlighting the complex dynamics of QCD. Zoomed in versions, including also a ratio of the data and theory predictions are shown for the bulk, perturbative, and transition regions in \Fig{fig:bulk}, \Fig{fig:pert}, and \Fig{fig:transitions}, respectively. Our measurement extends to $z\sim (1-z) \sim 10^{-5}\ll (\Lambda_{\text{QCD}}/Q)^2$, achieving for the first time a complete measurement of the energy correlator spectrum extending below the confinement scale in both the collinear and back-to-back limits.

\emph{{\color{alephblue}Theoretical Predictions.}} The calculation of the energy correlator spectrum at a level of precision comparable to our new measurement relies on tremendous progress in perturbative QFT, effective field theories, and lattice QCD \cite{Moult:2025nhu}. Here we provide a brief overview of our calculation, the most precise ever achieved, highlighting technical ingredients which enabled us to achieve this precision. A more detailed description of the calculation is provided in \cite{EEC:forthcoming}. 

The calculation of the energy correlator distribution in QCD is challenging due to the fact that it requires theoretical control of QCD dynamics over three orders of magnitude. To achieve this daunting task, we use effective field theory techniques to exploit the separation of scales: the energy correlator is a complicated multi-scale function, depending on the scales  $Q^2$, $Q^2(1-z)$, $Q^2 z$, and $\Lambda_{\text{QCD}}^2$. At LEP energies,  a large range of the EEC distribution has $Q^2\,,~Q^2(1-z)\,,~Q^2 z \gg \Lambda_{\text{QCD}}$. In these regimes we can derive factorization theorems, which rigorously separate universal long distance non-perturbative contributions from short distance contributions that can be computed in perturbation theory.   These factorization theorems are proven and formulated using soft-collinear effective theory (SCET) \cite{Bauer:2000ew,Bauer:2000yr,Bauer:2001ct,Bauer:2001yt,Rothstein:2016bsq}, and depend on the kinematic configuration of the energy flow operators. 
Predictions from the different factorization theorems are then merged by matching the expansions to achieve our final complete prediction.  The incorporation of tracking information in factorization theorems for the energy correlators was developed in \cite{Chang:2013rca,Chang:2013iba,Chen:2020vvp,Jaarsma:2023ell,Chen:2022pdu,Chen:2022muj,Jaarsma:2022kdd,Li:2021zcf}. Here we provide a brief summary of the different factorization theorems used in our calculations. In all our factorization theorems, we incorporate the running of the coupling to (up to) five loops \cite{Baikov:2016tgj} (as dictated by the underlying factorization theorem), and the evolution of the track function moments to three loops \cite{Jaarsma:2022kdd,Mitov:2006ic,Chen:2020uvt}.

In the bulk of the energy correlator, $z\sim 1/2$, we have a two-scale problem, allowing an expansion in $Q \gg \Lambda_{\text{QCD}}$. The perturbative distribution is computed analytically at leading and next-to-leading order \cite{Dixon:2018qgp,Li:2021zcf,EEC:forthcoming}, and numerically using data from ColorfulNNLO \cite{DelDuca:2016ily,DelDuca:2016csb,Tulipant:2017ybb} at next-to-next-to-leading order (NNLO$_\text{FO}$). The leading non-perturbative correction is captured by a single universal non-perturbative number \cite{Korchemsky:1999kt,Lee:2006fn}, whose value has been extracted from experimental measurements of event shapes \cite{Abbate:2010xh,Schindler:2023cww}.

The factorization theorem for the back-to-back limit of the energy correlator, $Q^2 \gg Q^2(1-z) \gg \Lambda_{\text{QCD}}^2 $ was derived in \cite{Moult:2018jzp}, building on the seminal works of \cite{Collins:1981uk,Collins:1981va}. All matrix elements appearing in this factorization theorem are known to three-loop order in QCD \cite{Luo:2019hmp,Luo:2019bmw,Ebert:2020qef,Li:2016ctv,Baikov:2009bg,Lee:2010cga,Gehrmann:2010ue}. Their scale evolution is governed by the so-called rapidity anomalous dimension, which is known to four-loop order \cite{Duhr:2022yyp,Moult:2022xzt}, and the cusp anomalous dimension, which is known to four loop \cite{Henn:2019swt}, and approximate five loop \cite{Herzog:2018kwj} order. Combining this tremendous combination of perturbative ingredients allows us to achieve next-to-next-to-next-to-next-to-leading logarithmic (NNNNLL$_{\text{b2b}}$) order. Extending earlier precision perturbative treatments of the back-to-back region of the energy correlator \cite{Moult:2018jzp,Duhr:2022yyp,Ebert:2020sfi}, we additionally incorporate the leading non-perturbative correction, which is described by the same universal constant as in the bulk of the distribution \cite{Dokshitzer:1999sh,EEC:forthcoming}, and the leading non-perturbative correction to the rapidity anomalous dimension extracted from recent lattice calculations \cite{Avkhadiev:2024mgd,Avkhadiev:2023poz,Shanahan:2021tst,Shanahan:2020zxr,Shanahan:2019zcq}.

The factorization theorem for the collinear limit of the energy correlator, $Q^2 \gg Q^2 z \gg \Lambda_{\text{QCD}}^2 $, was derived in \cite{Dixon:2019uzg}. All matrix elements appearing in the factorization are known to two loops \cite{Dixon:2019uzg,Mitov:2006wy}, and the relevant anomalous dimensions are known to three-loops \cite{Mitov:2006ic,Chen:2020uvt}.
This allows us to achieve next-to-next-to-leading logarithmic (NNLL$_{\text{col}}$) order. We additionally incorporate the leading non-perturbative correction, which is again given by the same universal constant as in the bulk of the distribution \cite{Korchemsky:1999kt,Chen:2024nyc,Lee:2024esz}.

Combining these ingredients, we are able to achieve the most precise calculation ever performed of the energy correlator observable over its full range, and furthermore, extend it to a calculation on tracks. Our final prediction is shown in \Fig{fig:plot_overall}, and in the zoomed versions in \Fig{fig:bulk},  \Fig{fig:pert}, and \Fig{fig:transitions}. As seen in the ratios between data and theory, this amounts to agreement at $\lesssim 5\%$ throughout much of the perturbative regime of the spectrum, with particularly good agreement in the perturbative back-to-back region, due to the extremely high perturbative order achieved there. This provides a beautiful illustration of the impact of the high-loop perturbative results in QFT on phenomenology. In our predictions we use the world average value of the strong coupling constant, $\alpha_s=0.118$ \cite{Huston:2023ofk}. We therefore emphasize that we do not have any free parameters in our calculation. Having illustrated the precision achievable with archival ALEPH data, and the precision of the theory calculation achievable for the EEC on tracks, we have opened the door to a program to extract precision constants of QCD from this data.

\emph{{\color{alephblue}Emergent States of QCD Come Into Focus.}} The unique aspect of our measurement of the EEC, \Fig{fig:plot_overall}, is that we have achieved for the first time a measurement of the spectrum covering three orders of magnitude in scale, all the way from partons, $Q^2z\,,~Q^2(1-z)\gg \Lambda_{\text{QCD}}^2$ to hadrons $Q^2z\,,~Q^2(1-z)\ll \Lambda_{\text{QCD}}^2$, including a precise resolution of the emergent states appearing in the transitions between these degrees of freedom in both the collinear and back-to-back limit. Here we describe the physics of \Fig{fig:plot_overall}, namely the EEC from partons to hadrons. 

We wish to emphasize that while the description of some regions of the EEC is under precise quantitative control, the description of other regions, in particular those involving transitions between distinct emergent states, is not. It is in these regions that our new data is arguably most valuable, and by attempting to associate regions of the EEC with precise field theoretic phenomena, we hope to take a first step towards sharpening our understanding of the phenomena controlling these regions.

The expectation value of the EEC is expressed as $\Tr[\rho_J \mathcal{E}(n_1) \mathcal{E}(n_2)]$, where $\rho_J=|\Psi_J\rangle \langle \Psi_J |$ is the density matrix for the state produced by acting with the current $J$ on the QCD vacuum. The effective description of this state depends strongly on the scale at which it is being probed by the EEC measurement, which is set by the kinematics of the detectors.

In the bulk of the distribution, $z\sim 1/2$, the current probes the QCD vacuum at the scale $Q\gg \Lambda_{\text{QCD}}$, producing neutral multi-parton quark and gluon states. In this regime we can organize the states into Fock space states of partons, with higher Fock space states suppressed by powers of the coupling. To NNLO$_{\text{FO}}$, we include the states $|q\bar q\rangle$, $|q\bar q g\rangle$, $|q\bar q g g\rangle$, $|q\bar q q\bar q\rangle$, $|q\bar q q\bar qg\rangle$, $|q\bar q g g g\rangle$. The energy correlator evaluated in these states has no particular symmetry properties, and is a complicated function of the angle. This regime is labeled in the center of \Fig{fig:plot_overall}, and a zoomed in version of this region is shown in \Fig{fig:bulk}. Excellent agreement between theory and data is observed. Focusing only on this central region, as was done in early measurements, hides the remarkable dynamics at small and large angles.

As we move the detectors towards the collinear limit, we can decompose the state into emergent degrees of freedom with definite quantum numbers under the Lorentz group, so called ``light-ray density" or ``light-ray quasi-particle" states. Additionally, we can decompose the product of detector operators into detectors of definite quantum numbers using the light-ray OPE \cite{Hofman:2008ar,Kologlu:2019mfz,Chang:2020qpj,Chen:2020adz,Chen:2021gdk,Chen:2023zzh}
\begin{align}
\mathcal{E}(\vec{n}_1)\mathcal{E}(\vec{n}_2) &= \sum_i \, (n_1\cdot n_2)^{\frac{\tau_i-4}{2}} C_i \mathbb{O}_{i}^{[J=3]} (\vec{n}_2) \,.
\label{eq:lightray_OPE}
\end{align}
Here $\mathbb{O}_{i}^{[J=3]}$ denote spin-3 light-ray operators, and $C_i$ their OPE coefficients. 
This expansion is organized as an expansion in the twist, $\tau$, of the contributing detectors or states. These detectors appearing in the OPE are dual to the light-ray states, acting to detect the light-ray quasi-particle states.
As $z\to 0$, but keeping $Q^2 z \gg \Lambda^2_\text{QCD}$ to remain in the perturbative regime, the measurement of the EEC isolates the twist-2 spin-3 light-ray states in QCD, of which there are a single quark and a single gluon state. 
At weak coupling, these states can be thought of as weakly dressed single particle states, as illustrated in \Fig{fig:pert}.
In this regime, the energy correlator reduces to a simple power law scaling, $1/z^{1-\gamma(3)}$, with a critical exponent governed by the (eigenvalues of the) anomalous dimensions of these light-ray states. The dressing only weakly modifies the scaling, as compared to the scaling associated with free states, namely $1/z$.
This enables the direct measurement of the non-integer scaling dimensions of an emergent, interacting quark and gluon state through precision measurements of correlations of \emph{hadrons}. It is also powerful, since the description of the correlator in this limit requires just the knowledge of the scaling anomalous dimensions, which are known to high precision. We can think of this scaling behavior as a ``standard candle" behavior of energy flux in QFTs, allowing us to precisely extract the value of their couplings from collider measurements. This scaling behavior is clearly visible in the indicated region in \Fig{fig:plot_overall}, and in the zoomed version of the small-$z$ region in Fig. \ref{fig:pert}. We see an excellent agreement, at the few percent level, between our calculations and experimental data.

As we move the detectors far apart, towards the $z\to 1$ (back-to-back) limit, the EEC exhibits a particularly interesting behavior in QCD. This arises due to the presence of a conserved gauge flux connecting the two energetic colored partons emitted from the QCD vacuum in the UV.  As compared to the collinear limit, where interactions only weakly dress the free parton states, we will see that the presence of a conserved flux will lead to a drastic modification of the behavior of the correlator in the back-to-back limit. Such a behavior can be understood either from an OPE perspective, arising from the logarithmic growth of anomalous dimensions in gauge theories, or from the perspective of emergent flux tube states: both manifestations of the presence of a conserved gauge flux. Both these perspectives lead to simplified descriptions of the EEC in the back-to-back limit enabling its precision calculation.

While the collinear limit of the EEC is sensitive to the twist-2 light-ray states and their anomalous dimensions at low spin ($J=3$), the back-to-back limit of the energy correlator is sensitive to the high spin states (the light-cone limit of the four-point correlator \cite{Korchemsky:2019nzm,Chen:2023wah}). This provides access to one of the unique features of a gauge theory, namely the logarithmic growth of the twist-2 anomalous dimensions at large spin \cite{Korchemsky:1988si,Korchemsky:1992xv,Gubser:2002tv,Belitsky:2006en}
\begin{align}\label{eq:large_spin_def}
\hspace{-0.15cm}\tau= 2+\Gamma_{\text{cusp}}(\alpha_s) (\log S+\gamma_E) + B_\delta(\alpha_s) +\mathcal{O}(1/S)\,,
\end{align}
governed by the cusp anomalous dimension, $\Gamma_{\text{cusp}}(\alpha_s)$, and virtual anomalous dimension $B_\delta(\alpha_s)$.
This logarithmic growth significantly dresses the large spin light-ray states, modifying them from free parton states, and giving them an effective size of $\mathcal{O}(\log(S))$.
Such a logarithmic growth of the anomalous dimensions is associated with the presence of a conserved flux \cite{Alday:2007mf}. Due to the presence of quarks in the fundamental representation, which can screen the flux in QCD, this is only true in the UV. This behavior is modified in the IR where the flux tube can freely break, as we will discuss in the next section. In theories without a conserved flux, we expect free behavior at large spin \cite{Komargodski:2012ek}. 

Using the simplified behavior of the correlator in the large spin limit, we can derive an all orders expression in the $z\to 1$ limit (keeping $Q^2 (1-z) \gg \Lambda^2_{\text{QCD}}$ to remain perturbative). The complete result in QCD can be found in \cite{EEC:forthcoming}, here we give a simplified form for a conformal gauge theory \cite{Belitsky:2013ofa,Korchemsky:2019nzm},
\begin{align}
\label{eq:b2b_factor}
&\text{EEC}_{z\to 1}=\frac{H(\alpha_s)}{8 (1-z)} \int\limits_0^\infty \df b\, b\, J_0(b) \\
&\exp \biggl[ -\frac{1}{2}\Gamma_{\text{cusp}}(\alpha_s)\ln^2 \Bigl( \frac{e^{2\gamma_E}\, b^2}{4(1-z)} \Bigr)+2B_\delta(\alpha_s) \ln \Bigl( \frac{e^{2\gamma_E}\, b^2}{4(1-z)} \Bigr)     \biggr]\,, \nonumber
\end{align}
expressed in terms of the leading coefficients of the large spin twist-2 anomalous dimension (Eq. \ref{eq:large_spin_def}), and a perturbatively computable matching coefficient $H(\alpha_s)$.
The integral represents a sum over the many high spin states that contribute in this limit.  As with the collinear limit, this provides a ``standard candle" scaling which we can compute to extremely high precision, enabling us to achieve a precise description of the energy correlator in this limit.

Much like the collinear limit, this simplified form of the back-to-back limit of the energy correlator in gauge theories should be interpreted as the appearance of emergent degrees of freedom. Indeed, in the back-to-back limit, the expectation value of the correlator $\langle J(x) \mathcal{E}(n_1) \mathcal{E}(n_2) J(0) \rangle$ is dominated by a classical saddle point describing a flux tube state \cite{Alday:2007mf}, whose action gives Eq. \ref{eq:b2b_factor}. To see this, we imagine the two energetic particles which excite the detector as framing Wilson lines (defects) in the QCD vacuum \cite{Alday:2010zy,Chen:2025ffl}. These are illustrated by the green lines in \Fig{fig:pert}. For light-like defects in gauge theories, the expectation value of the defect is computed by a saddle, which is dominated by a flux tube configuration \cite{Alday:2007mf}, whose energy density is described by the cusp anomalous dimension. We can therefore interpret our measurement as directly probing the action of this flux tube state for different configurations of the Wilson lines, set by the kinematics of the detectors. We find it remarkable that we can see the imprint of this perturbative flux tube state in the distribution of final state hadrons. 

An interesting feature of the result in Eq. \ref{eq:b2b_factor}, originally observed in \cite{Parisi:1979se}, is that it asymptotes to a constant as $z\to 1$.  As compared to the weak dressing of the light-ray states in the collinear limit, the states in the back-to-back limit are strongly modified from their free field value, changing the behavior by an integer power, from $1/1-z$ to $1$. This is consistent both with the intuition of the logarithmically enhanced dressing at large $S$, effectively growing the point particle states to quasi-particles of size $\mathcal{O}(\log(S))$, and from the perspective of a flux-tube state, which is a strong modification of free parton states.

The back-to-back limit of the energy correlator is highlighted in \Fig{fig:plot_overall}, and a zoomed in view is shown in \Fig{fig:pert}. We observe exceptional agreement, at the level of a few percent, between our calculation and our experimental measurement.  This agreement strongly highlights the tremendous progress in perturbative QFT, which has enabled a description of QCD energy flux at this remarkable accuracy. Our measurement also provides a clear experimental illustration that this perturbative flux tube state describes the asymptotics of the four-point function. It would be fascinating to further develop the relation between the back-to-back limit of the energy correlator and flux-tube states, to understand if one can directly image flux-tube states at electron-positron colliders, or to develop additional ways in which properties of flux tubes can be measured in colliders.

\emph{{\color{alephblue}Imaging the Confinement Transition.}}  As we lose the hierarchy $z Q^2 \gg \Lambda_{\text{QCD}}^2$, or $(1-z) Q^2 \gg \Lambda_{\text{QCD}}^2$, we lose the ability to compute the energy correlator using any known technique. However, one of the most remarkable features of our measurement is that it extends through, and even an order of magnitude beyond, the confinement transition, into the regime $z Q^2 \ll \Lambda_{\text{QCD}}^2$, and $(1-z) Q^2 \ll \Lambda_{\text{QCD}}^2$.  Since QCD is a gapped theory, there are no interactions below the confinement scale, and we expect the correlator to asymptote to a flat distribution characteristic of a uniform gas of hadrons.  Remarkably, in \Fig{fig:plot_overall}, as well as the zoomed versions of this limit in \Fig{fig:transitions}, we see that this is indeed the case. This is the first measurement illustrating this behavior in the back-to-back limit.

In \Fig{fig:transitions}, we are able to directly measure the transition between a scaling region where we have a precise description in terms of perturbative light-ray or flux-tube states, and a free hadron region. While we cannot describe these transitions, as illustrated by the breakdown of our first principles calculations, our theoretical understanding of the degrees of freedom on either side of the transition in each case,  allows us to give the transitions a precise theoretical interpretation.

In the back-to-back limit, for $(1-z)\gg \Lambda_{\text{QCD}}^2/Q^2$, we have an approximately conserved gauge flux, and our correlator is described by a perturbative flux tube state. 
The fact that Eq. \ref{eq:b2b_factor} asymptotes to a constant (assuming fixed coupling), matching onto the free hadron result, has important implications for understanding the impact of confinement on the energy correlator distribution in the back-to-back limit.
The value of $z$ at which the correlator asymptotes to a flat distribution in the back-to-back limit is not set by confinement, but rather the properties of the flux tube.  Indeed, this turnover occurs in a conformal gauge theory. In QCD, there are two modifications to this picture. First, due to the running coupling, the anomalous dimensions themselves become non-perturbative in this regime.  We are able to slightly extend the validity of our calculations using calculations from lattice QCD \cite{Avkhadiev:2024mgd,Avkhadiev:2023poz,Shanahan:2021tst,Shanahan:2020zxr,Shanahan:2019zcq}. As can be seen in \Fig{fig:transitions}, while our calculation has the required turnover to a flat distribution, we lose agreement with the data, highlighting that our new data in this regime will be extremely useful. 
Second, in the deep IR, $(1-z)\to \Lambda_{\text{QCD}}^2/Q^2$, the QCD flux tube can freely break into hadrons. This occurs smoothly in the energy correlator distribution, since the perturbative prediction has already transitioned to a uniform distribution. Our new data provides the most precise data in this regime of flux tube breaking, and we hope that it can be used to sharpen our understanding of the QCD flux tube.

In the collinear limit, for $z\gg \Lambda_{\text{QCD}}^2/Q^2$, the energy correlator exhibits a power law scaling describing the dynamics of the twist-2 spin-3 light-ray states. This scaling is universal in any approximately conformal theory, and in a genuine conformal theory, persists all the way to $z=0$. 
In QCD, which exhibits both a running coupling, and confinement, a much more complicated behavior occurs: as the coupling increases, the light-ray state becomes increasingly strongly dressed, departing significantly from its free parton state. As $z\to \Lambda_{\text{QCD}}^2/Q^2$, the scaling is broken by the generation of a mass scale, and the light-ray state hadronizes into color singlet states, causing a rapid transition to a uniform scaling behavior associated with free hadrons. As opposed to the transition in the back-to-back limit, which is captured in perturbation theory, the transition in the collinear limit is not captured in resummed perturbation theory, rather it is associated with the generation of a mass scale, $\Lambda_{\text{QCD}}$, and the $z$-value of the transition is directly set by confinement. Although we cannot compute this transition, understanding the states on either side of the transition we can precisely understand its interpretation, namely it is the overlap of a twist-2 light-ray state with a two-hadron state \cite{Chang:2025kgq}, as illustrated in \Fig{fig:transitions}.  This interpretation makes manifest why the collinear transition cannot be captured in perturbation theory, unlike the back-to-back transition. In the language of QCD factorization, the collinear transition is described by di-hadron fragmentation functions \cite{Lee:2025okn,Herrmann:2025fqy,Kang:2025zto}, and precision measurements of this region can be used to extract these di-hadron fragmentation functions. Therefore we see that the transition in the collinear limit of the EEC allows us to image another manifestation of confinement, quite distinct from the breaking of the flux tube in the back-to-back limit.

We hope that the quality of the data we have obtained in these transition regions, combined with the sharp theoretical interpretation of the transitions that we have provided, will enable improvements in our understanding of confinement.

\emph{{\color{alephblue}Conclusions and Outlook.}} In this \emph{Letter}, we unlocked thirty-year-old ALEPH  data to deliver the highest angular resolution map of QCD energy flow.
Using this data we performed a high-angular resolution measurement of the two-point correlation of energy flux, providing a clean measurement of the dynamics of QCD over three orders of magnitude in scale.
Our new data, combined with a detailed theoretical understanding of the physics of the energy correlator, enables both precision tests of QCD in well understood regions of the EEC, as well as new measurements of field theoretic phenomenon in poorly understood regions. 
This combination will prove invaluable for advancing our understanding of QCD.

This \emph{Letter} demonstrates the potential for archival datasets to create a broad physics program that actively contributes to the forefront of exploration. The precision illustrated in this paper suggests the possibility for precision measurements of the strong coupling constant, hopefully enabling the resolution of long-standing tensions between extractions of the strong coupling constant from $e^+e^-$ event shapes \cite{Benitez:2025vsp,Benitez:2024nav,Benitez-Rathgeb:2024ylc,Hoang:2014wka,Hoang:2015hka,Abbate:2010xh}, and the world average. Beyond this, the quality of the data in the non-perturbative transition regions of the energy correlator will prove invaluable for the study of universal non-perturbative parameters, such as di-hadron fragmentation functions, and the Collins-Soper kernel. Many other measurements of energy correlator observables, such as higher-point correlation functions \cite{Chen:2019bpb,Yan:2022cye,Yang:2022tgm,Chicherin:2024ifn,Ma:2025qtx} or azimuthal asymmetries \cite{Kang:2023big}, will further explore this dataset, and we envision equally transformative gains in resolution through the use of tracks in those cases.

More broadly, we are excited by the interplay between archival data, and data from the LHC and future electron-ion collider (EIC).  The asymptotic limits of the energy correlator measured in this paper are closely related to the $Z$ and Higgs boson transverse momentum distributions, some of the most precisely measured distributions at the LHC \cite{ATLAS:2014alx,ATLAS:2015iiu,ATLAS:2019zci,ATLAS:2023lsr,CMS:2011wyd,CMS:2016mwa,CMS:2019raw,LHCb:2015mad,LHCb:2016fbk}. Having brought archival LEP data to the level of precision required to contribute to the precision LHC program, we anticipate a bright future interplay between archival and modern data for challenging our understanding of QCD and the Standard Model.

\emph{{\color{alephblue}Acknowledgements.}}---The authors thank Roberto Tenchini, Guenther Dissertori and Paolo Azzurri from the ALEPH Collaboration for their useful comments and suggestions on the use of ALEPH data. We thank Stephen Ellis, Gregory Korchemsky, George Sterman, Lance Dixon, David Simmons Duffin, Juan Maldacena, Hao Chen, Xiaoyuan Zhang, Andrew Tamis, Kyle Lee, Jesse Thaler, Victor Gorbenko, Pedro Vieira, Phiala Shanahan, Gherardo Vita, and Sasha Zhiboedov for many useful discussions, collaborations, and encouragement to achieve precision measurements of energy correlators in electron-positron colliders.
We thank Tilman Plehn for coining the term ``recycling frontier" at the BOOST conference in 2025.
We thank all the members of the ``$e^+e^-$ alliance" for their tremendous efforts to resurrect the LEP dataset, ultimately enabling the studies presented in this paper.
Y.L. is supported by funding from the European Research Council (ERC) under the European Union’s Horizon 2022 Research and Innovation Program (ERC Advanced Grant agreement No.101097780, EFT4jets).
I.M. is supported by the U.S. Department of Energy (DOE) Early Career Award DE-SC0025581, and the Sloan Foundation. 
Y.C. and J.Z. are supported in part by the U.S. National Science Foundation under ID 2514008.
H.X.Z. is supported by the National Natural Science Foundation of China under contract No.~1242550.
H.B., Y-C.C., and Y-J.L. are supported by the U.S. Department of Energy, Office of Science, Office of Nuclear Physics under grant Contract Number DE-SC0011088.
\bibliography{EEC_ref.bib}{}
\bibliographystyle{apsrev4-1}
\newpage
\onecolumngrid
\newpage

\end{document}